\documentclass[%
preprint,
superscriptaddress,
 amsmath,amssymb,
floatfix,
]{revtex4-1}
\usepackage{graphicx}  
\usepackage{ulem}      
\usepackage{lineno}
\usepackage[colorlinks]{hyperref}   

\begin{document}


\title{Plasmon-induced efficient hot carrier generation in graphene on gold ultrathin film with periodic array of holes: Ultrafast pump-probe spectroscopy}

\author{Gyan Prakash}
 \affiliation{Department of Physics, Indian Institute of Science, Bangalore 560 012, India.}
\affiliation{Center for Ultrafast Laser Application, Indian Institute of Science, Bangalore 560 012,
India.
}%

\author{Rajesh Kumar Srivastava}%
\affiliation{Department of Physics, Indian Institute of Science, Bangalore 560 012, India.
}%

\author{Satyendra Nath Gupta}
\affiliation{%
Department of Physics, Indian Institute of Science, Bangalore 560 012, India.
}%

\author{A. K. Sood}
\affiliation{%
Department of Physics, Indian Institute of Science, Bangalore 560 012, India.
}%

\affiliation{Center for Ultrafast Laser Application, Indian Institute of Science, Bangalore 560 012,
India.
}%


\begin{abstract}
Using ultrafast pump-probe reflectivity with 3.1 eV pump and coherent white light probe (1.1 to 2.6 eV), we show that graphene on gold nanostructures exhibits a strong coupling to the plasmonic resonances of the ordered lattice hole array, thus injecting a high density of hot carriers in graphene through plasmons. The system being studied is  single-layer graphene on  ultrathin film of gold with periodic arrangements of  holes showing anomalous transmission. A comparison is made with gold film with and without hole array. By selectively probing transient carrier dynamics in the spectral regions corresponding to plasmonic resonances, we show efficient plasmon induced hot carrier generation in graphene.  We also show that due to high electromagnetic field intensities at the edge of the sub-micron holes, fast decay time (10-100 fs) and short decay length (1 nm) of plasmons, a highly confined density of hot carriers (very close to edge of the holes) is generated by Landau damping of plasmons within the holey gold film. A contribution to transient decay dynamics due to diffusion of initial non-uniform distribution of hot carriers away from the hole edges is observed.   Our results are important for future applications of novel hot carrier device concepts where hot carriers with tunable energy can be generated in different graphene regions connected seamlessly. 
\end{abstract}

\maketitle

\section{Introduction}
Surface plasmon polaritons (SPPs) due to their inherent property of nanoscale confinement and localization, smaller than the interacting light wavelength, show strong light-matter interaction. This leads plasmonics to a wide variety of applications from efficient harvesting of solar energy to nanoscale optical devices \cite{genet2007light, barnes2003surface, brongersma2015plasmon}. Efficient generation and transfer of hot electron-hole  pairs has been at the heart of efficient conversion of solar energy in photovoltaic and photocatalytic devices \cite{clavero2014plasmon}. Plasmons in nanostructures decay  through Landau damping (with damping time of few tens of fs \cite{moskovits2015case, link1999size, sonnichsen2002drastic, link2000shape}) to generate energetic electrons, with energies larger than those of the thermal excitation  (known as the hot electrons) at ambient temperatures,  \cite{tsu1967landau, wu2015efficient, sundararaman2014theoretical, ropers2005femtosecond, boriskina2013plasmonic, shahbazyan2016landau}.  Hot carriers can be accessed through  plasmon induced hot-electron charge transfer  into adjacent interface material  \cite{wu2015efficient, govorov2013theory, sonnichsen2002drastic, hartland2011optical}.  If a high conductance pathway is available, the plasmon generated hot electrons can rapidly conduct over long distances, which might also discourage the electrons once transferred from  returning to the plasmonic material. In graphene, electrons move with high velocities ($v_{_F}$=10$^6$ m/s) and behave like massless Fermions, making it the most desirable material for the high-speed electronics and optoelectronics applications. A substantial challenge to the performance of most of the graphene-based optoelectronic devices \cite{novoselov2004electric, zhang2005experimental, grigorenko2012graphene}  comes from the weak absorption of light (only 2.3\% at normal incidence) in graphene. Thus, efficient hot electron injection in graphene through plasmons gains valuable importance. Graphene in a hybrid plasmonic  structure are promising elements for high-speed electrically controllable optical and optoelectronic devices. Two dominant mechanisms to induce hot carriers in graphene are (i) direct hot electron transfer by Landau damping of plasmons in Au nanostructures, as seen in reduced graphene oxide coated gold nanoparticles \cite{kumar2016ultrafast} and graphene covered gold nanorods \cite{hoggard2013using} and (ii) generation of carriers by near field enhanced electric field as reported in graphene covered nanodisks \cite{gilbertson2015plasmon}.  \textcolor{black}{Ultrafast pump-probe experiments have played important role in identifying the hot carrier transfer mechanism and dynamics \citep{wu2015efficient, shan2019direct, shan2019electron}.}

  Ever since the phenomenon of extraordinary optical transmission (EOT) was demonstrated \cite{ebbesen1998extraordinary, genet2007light}, plasmonic structures with periodic array of holes in metal films owing to  high refractive index sensitivity (RIS) and tunable plasmonic properties have found  wide applications from optical elements to chemical and biological sensors \cite{psaltis2006developing,  cetin2014handheld, escobedo2013chip}. Graphene on plasmonic nanostructures of noble metals is the simplest and effective hybrid structure for seamless integration of graphene with plasmons. Recent experiments have shown a strong light-graphene interaction enhanced by plasmons  of array of holes and nanovoids  in  gold \cite{zhu2013enhanced, reckinger2013graphene}.  Of particular interest are the nanovoids with protruding edges which show strong graphene-plasmon interaction due to relatively large interaction area between graphene and the strong electric field of plasmons at the voids or rims. Despite having the possibility of more strong interaction of graphene on 3D hole arrays due to large interaction area (similar to the case of nanovoids) and enhanced field intensities (because of matched plasmon energy between the two sides), this has not been explored.   Most of the studies of graphene-covered periodic hole/ nanovoid arrays  so far  have focused on the enhanced  environmental sensitivity of plasmonic resonances \cite{reckinger2013graphene, zhu2013enhanced, hao2012surface, zhang2015gold, genslein2016graphene, mahigir2017plasmonic, zhao2017toward}.  The ultrafast interplay of dynamics of carriers and plasmons  in graphene-covered metal hole array is completely unexplored.

Here, we show that a high density of hot-electrons can be generated in graphene, through a strong interaction in a hybrid plasmonic structure of graphene with 3D gold hole array. Using ultrafast pump pulses  ($\sim $80 fs) of energy 3.1 eV we photoexcite carriers in graphene-gold hole array hybrid system and  probe their relaxation dynamics with  coherent white light in the spectral window of 1.1 eV to 2.6 eV. Notably, pump-induced reflectivity shows significant signatures in the  spectral window corresponding to EOT resonances of gold hole array  originating from the carrier dynamics in graphene.  A comparative study of the graphene on gold film with and without hole array confirms the highly efficient direct plasmon induced hot carrier generation in graphene.

\section{\label{sec:ExpDetails}  Experimental and simulation details}

\begin{figure*}
\centering
\includegraphics[width=0.8\linewidth]{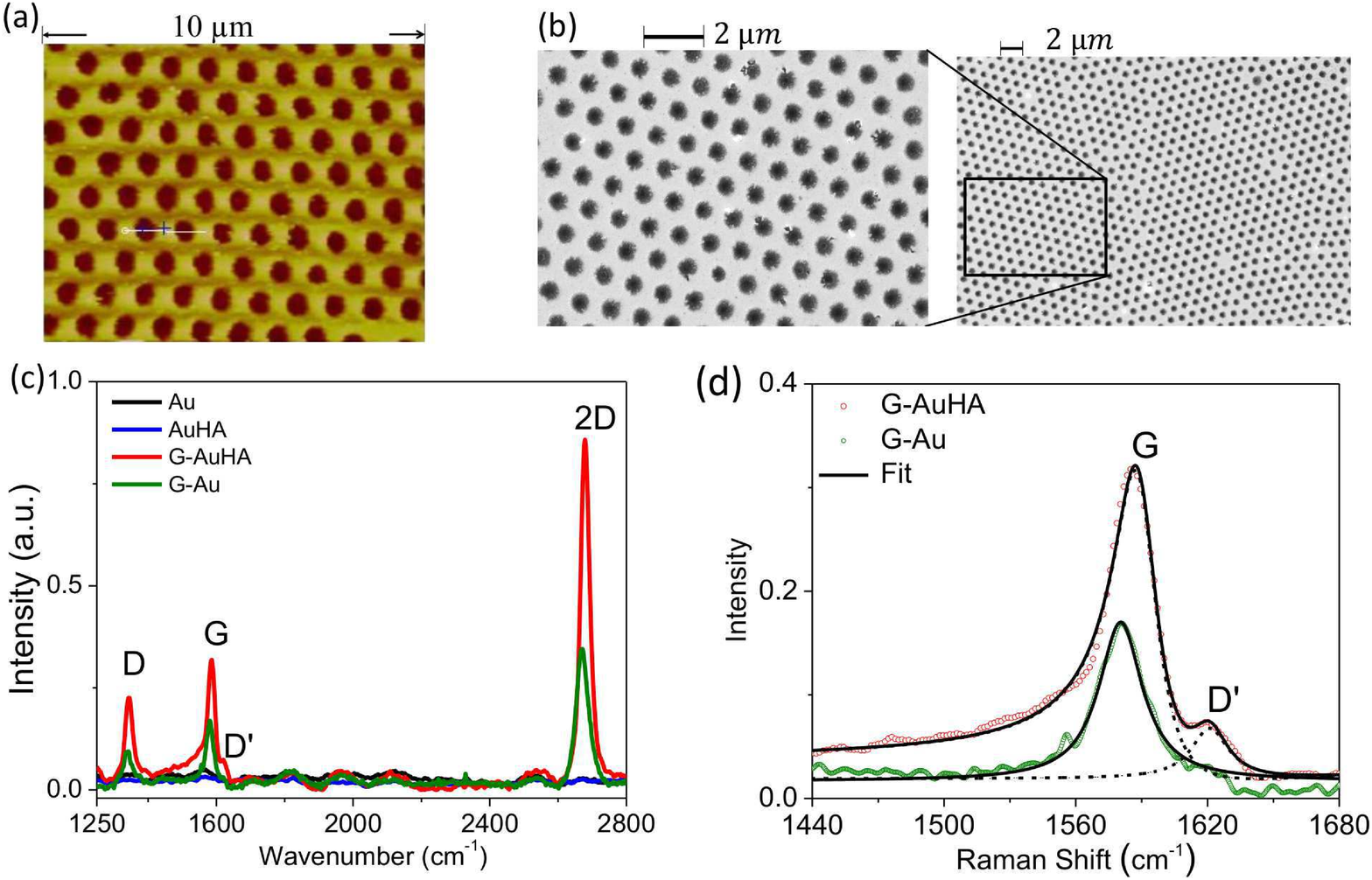}
\caption{(Color Online) (a) AFM and (b)SEM images of ordered triangular lattice array of holes. (c) Raman spectrum of single layer graphene transferred on AuHA and Au. (d)Asymmetric lineshape of G band in G-AuHA compared with G band of G-Au. The solid line in G-Au are fit to the experimental data with a Lorentzian whereas in G-AuHA are with Brat-Wigner-Fano (BWF) lineshape.}  
\label{fig:Samp&Raman}
\end{figure*}

Gold hole array was fabricated on a glass substrate using colloidal lithography technique \cite{fredriksson2007hole}. The hole array was prepared by spin coating solution of polystyrene spheres of 1 $\mu$m mean diameter on a glass substrate for a hexagonal close-packed colloidal monolayer. The polystyrene sphere diameter was reduced to about 480 nm with reactive ion etching using O$_2$ plasma. Subsequently, 50 nm Au with 5 nm Ti adhesion layer was deposited on the microspheres by sputtering. The  removal of polystyrene spheres by ultrasonication gives holes of 490 $\pm$30 nm diameter in gold film.   The atomic force microscopy (AFM) and scanning electron microscopy (SEM) were used to characterize the morphology of fabricated hole array (see Fig.  \ref{fig:Samp&Raman}(a) and (b)). The array of holes with an average diameter of 490 nm are arranged in a triangular lattice of  1 $\mu$m periodicity.

Single-layer graphene samples were grown on a 25 $\mu$m thick copper foil using chemical vapor deposition technique and transferred on gold films with and without hole array by conventional technique \citep{kar2014tuning,rao2009graphene}.  The PMMA on the graphene was removed by immersing the substrate in acetone at 60 $^0$C, followed by annealing for 3h at 500 $^0$C under hydrogen and argon atmosphere. For a reference, Au film of 50 nm thickness without hole array with Ti adhesion layer was deposited on a glass substrate. We have studied four samples: (i) 50 nm gold film with hole array (AuHA), (ii) graphene covered gold hole array (G-AuHA), (iii) 50 nm gold film (Au) and (iv) graphene covered 50 nm gold film (G-Au).

Fig.  \ref{fig:Samp&Raman}(c) shows Raman spectrum of G-AuHA and G-Au with excitation wavelength of 532 nm. The peak positions of G mode (1590 cm$^{-1}$), 2D mode (2680 cm$^{-1}$ with width of $\sim$31 cm$^{-1}$)  as well as their intensity ratio  confirm the single layer graphene on G-AuHA \citep{das2008monitoring}. The D band intensity with respect to the G-band (I$_{D}$/I$_G$ $\sim$ 0.68) shows presence of disorder in the single layer graphene. The D' band (1621 cm$^{-1}$) associated with the disorder activated  longitudinal optical phonons near $\Gamma$-point is also seen in G-AuHA.  As work function of graphene ($\sim$  4.56 eV \citep{yan2012determination}) is lower than that of gold ($\sim$4.83 eV \citep{anderson1959work}), graphene on gold film gets p-doped. Further  CVD grown graphene gets unintentionally p-doped due to charge transfer from atmospheric H$_2$O/O$_2$ molecules.   The intensity ratio I(2D)/I(G) is $\sim$ 2.7 and 2.0 for graphene on AuHA and graphene on Au, respectively. From the I(2D)/I(G) ratio, we find that graphene on Au is more p-doped than graphene on AuHA with respective Fermi energies at -352 meV and -206 meV, respectively \citep{das2008monitoring}. This is commensurate with the fact that graphene is relatively  in more contact with gold in the case of G-Au than G-AuHA.  A enhancement of $\sim$2 in Raman intensity of D, G and 2D modes in G-AuHA in comparison to G-Au is due to the enhanced electric field originating from the nanostructure  \citep{zhu2013enhanced}.  The G band of G-AuHA shows asymmetric lineshape (Fig. \ref{fig:Samp&Raman}(d)).   We fit the G band with  Breit-Wigner-Fano (BWF) line shape \citep{cerdeira1973effect}: $I(\omega)=I_0[1+(\omega-\omega_{G})/q\Gamma]^2/[1+[(\omega-\omega_{G})/\Gamma]^2]$ (where $\Gamma$ is the linewidth, $\omega_G$ is the G band frequency, and 1/q is the interaction parameter between the phonon and electronic continuum). From the fit we obtain $\omega_G$=1590.0$\pm$0.3 cm$^{-1}$, $\Gamma$=12.1$\pm$0.3 cm$^{-1}$ and -1/q=0.23$\pm$0.01.  From the Lorentzian fit to the symmetric G band in G-Au the values are $\omega_G$=1580.0$\pm$0.1 cm$^{-1}$, $\Gamma$=12.0$\pm$0.2 cm$^{-1}$.  Similar values of 1/q as seen by us have been reported for metallic single-walled carbon nanotubes (SWNT) where lower frequency  (G band) transverse optical (TO) phonon component couples to the  1D-$\pi$ plasmons \citep{brown2001origin}.  Asymmetric lineshapes  have been observed in graphene as a manifestation of a Fano resonance originating from the interaction of G-phonons with the excitonic states formed in undoped graphene \citep{yoon2013fano}. The interaction parameter 1/q observed by us is about a factor of three greater than that observed in pristine graphene with Fermi energy at the charge neutrality point\citep{yoon2013fano}. We attribute the observed BWF lineshape to the interaction of G phonons  with the   \textcolor{black}{elctronic Raman background of} AuHA, implying  strong coupling of graphene with the plasmonic nanostructure.

Transient reflectivity measurements were done by exciting the  samples with 80 fs  pulses of central wavelength 400 nm (3.1 eV), obtained by frequency doubling 800 nm (1.57 eV ) derived from Ti: Sapphire amplifier (Spitfire from Spectra Physics Inc.) with a BBO crystal. Pump-induced reflectivity changes  were monitored for Au, G-Au, AuHA and G-AuHA with stable white light probe (generated by focusing a small fraction of 800 nm on a sapphire crystal) in  spectral window  of 1.1 eV to 2.6 eV. Pump beam with fluence of 138 $\mu$J/cm$^2$  (for Au, AuHA, and G-AuHA) and 201 $\mu$J/cm$^2$ (for G-Au)  with an incident angle $\sim 14^0$ were used.

The steady state transmission measurements were done using UV/VIS spectrophotometer (Lambda 950 Perkin Elmer). Optical transmission of the AuHA and G-AuHA is simulated using 3D full wavevector FDTD method using  software package (FDTD Solutions from M/s Lumerical Inc.). A plane polarized source with polarization along x-axis propagating in the z-direction normal to AuHA and G-AuHA is considered.  A transmission intensity monitor at the back side  and a field intensity monitor at the front surface of AuHA and G-AuHA were kept to record the simulated light transmission and the electric field intensity at the surface, respectively. For accounting the irregularities in the circular shape of holes and better agreement with the experimentally observed transmission, the simulated transmission was averaged over transmission spectra from slightly elliptical holes (keeping the sum of major and minor axis radii 500 nm) with the same periodicity.

\section{\label{sec:Result&Discuss} Results and discussion}
\begin{figure*}[h]
\centering
\includegraphics[width=0.64\linewidth]{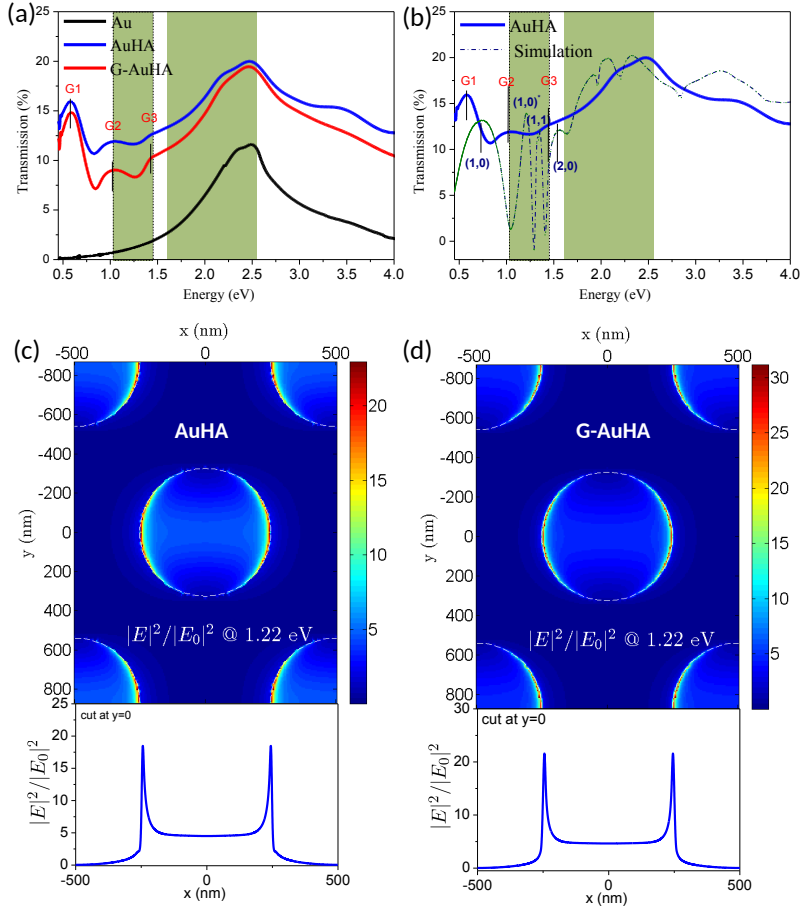}
\caption{(Color Online)(a)Transmission spectrum of  AuHA and G-AuHA showing extraordinay optical transmission (EOT) peaks due to coupling of light with the plasmonic resonances of ordered lattice hole array. Respective plasmonic mode indices are also indicated. For comparison transmission spectrum of gold film (Au) of same thickness is also included (the plots are not shifted vertically).  (b) FDTD simulated transmission spectrum showing  (1,0) AuHA/glass, (1,0)$^*$ Air/AuHA, (1,1) AuHA/Glass and (2,0) AuHA/Glass interface EOT peaks. For comparison, experimental AuHA transmission data in (a) is also included.
Green shaded regions refer to the spectral window covered by white light probe in transient differential reflectivity measurement experiment. (c) and (d) Electric field intensity profile  and plot at y=0 for EOT  (1,0)$^*$ Air/AuHA and (1,0)$^*$ Air/G-AuHA at 1.22 eV, respectively. } 

\label{fig:Transm&FDTD_E_prof_plot}
\end{figure*}

\begin{table*}[h]
\centering
\caption{\label{tab:table1} Electric field intensity ratio $R$ for EOT frequencies observed in FDTD simulated transmission spectrum of AuHA and G-AuHA.}
\begin{tabular}{c c c c c c}
Calculated EOT peak position & \hspace{1 cm} &Assignment & $R$ (AuHA)&  $R$ (G-AuHA) &  $\%$ Enhancement \\

\hline
0.83 eV  &G1  & (1,0) AuHA/Glass     & 369    &  543 &  47 $\%$ \\
1.22 eV  &G2  & (1,0)$^*$ Air/AuHA   & 562    &  869 &  55 $\%$\\
1.35 eV  &G2  & (1,1) AuHA/Glass     & 159    &  276 &  74 $\%$\\
1.55 eV  &G3  & (2,0) AuHA/Glass     & 7.8    &  8.7 &  11 $\%$\\

\hline
\end{tabular}
\vspace*{15pt}
\end{table*}
Fig.  \ref{fig:Transm&FDTD_E_prof_plot}(a) shows the measured transmission spectra of AuHA and G-AuHA showing extraordinary optical transmission (EOT) peaks due to  triangular lattice array of holes. For comparison, transmission spectrum of the gold film of same thickness is also shown. The transmission spectrum of AuHA and G-AuHA are not vertically shifted in Fig.  \ref{fig:Transm&FDTD_E_prof_plot}(a),  thus showing enhanced transmission with respect to Au.  Transmission spectra of AuHA and G-AuHA show three EOT transmission peaks whose positions are related to the lattice constant $a_0$ of the hole array,  dielectric constants of the surrounding medium ($\epsilon_d$) and the metal ($\epsilon_m$) and the Bragg resonance order (i, j) of the SPPs satisfying Bragg condition \citep{genet2007light}
\begin{equation}
\begin{aligned}
\nu_{max}^{(i,j)}=ca_0^{-1}\left[\left(4/3\right)\left( i^2+ij+j^2\right)\left(\epsilon_m^{-1} + \epsilon_d^{-1}\right)\right]^{1/2}
\end {aligned}
\end{equation}

We observe three EOT peaks in the transmission spectrum at 0.6 eV, 1.0 eV  and 1.4 eV,  marked as G1, G2 and G3 in Fig.  \ref{fig:Transm&FDTD_E_prof_plot}(a), respectively.  The observed EOT peak positions are in good agreement  with the  reports on similar gold hole array  films  \citep{ai2012novel, quint2014getting}. The simulated transmission spectrum for triangular lattice gold hole array of 500 nm diameter and 1000 nm periodicity on glass substrate is shown in Fig.  \ref{fig:Transm&FDTD_E_prof_plot}(b).  This guides us in assigning the EOT peaks as follows (Table I): the transmission peak  G1 is due to (1,0) AuHA/Glass Bragg resonance order, the broad EOT peak G2 is due to the overlap of (1,0)$^*$ Air/AuHA  and (1,1) AuHA/Glass resonance peaks and the G3 peak is attributed to (2,0) AuHA/Glass resonance peak. A slight blue shift of the simulated spectrum with respect to the measured one can be due to the irregularities in the long range periodicity of triangular lattice of the hole arrays which has not been considered in the simulation.
An overall enhanced transmission of $\sim$10 $\%$ in AuHA with respect to Au is observed in the spectral window from 1.6 eV to 4.0 eV where EOT peaks are not seen. G-AuHA due to absorption in graphene (2.3 $\%$) shows a slightly reduced transmission with respect to AuHA in the same spectral window. However, in the spectral window of 0.5 eV to 1.6 eV  at EOT peaks, due to coupling of graphene with the plasmonic modes, transmission is reduced by $\sim$ 3.5 $\%$ \citep{zhu2013enhanced, reckinger2013graphene}.

The simulations with graphene on top of the AuHA did not show any difference in the transmission peak positions. However, the electric field intensity is enhanced due to graphene by more than $\sim$55\%.  Fig.  \ref{fig:Transm&FDTD_E_prof_plot}(c-d)  show the simulated electric field intensity distribution profile on top surface AuHA and G-AuHA, respectively for the photon energy corresponding to  (1,0)$^*$ Air/AuHA and (1,0)$^*$ Air/G-AuHA  EOT resonance frequencies. The electric field intensity profiles show that the field intensity is high at the edge of the holes ( see electric field intensity plots at y=0 in Fig. \ref{fig:Transm&FDTD_E_prof_plot}(c-d)). Table (I) summarizes calculated intensity ratio $R=|E|^2_{edge}/|E|^2_{mid}$, where $|E|^2_{edge}$  and $|E|^2_{mid}$ are the electric field intensities at the edge and at a mid point  between the two adjacent holes. It can be seen that the electric field intensities at the hole edge are higher at G1 and G2 EOT frequencies. The (1,0)$^*$ Air/AuHA shows  greater $|E|^2_{edge}/|E|^2_{mid}$ value than the (1,0) AuHA/Glass as the SPP is on the Air/AuHA interface, where the near field monitor is placed.  The corresponding electric field intensities are enhanced when  AuHA is covered with graphene (see Table (I)).  We will see later that the intense electric field intensities at the edge of the holes play an important role in hot carrier generation in graphene.

\begin{figure*}[h]
\centering
\includegraphics[width=0.8\linewidth]{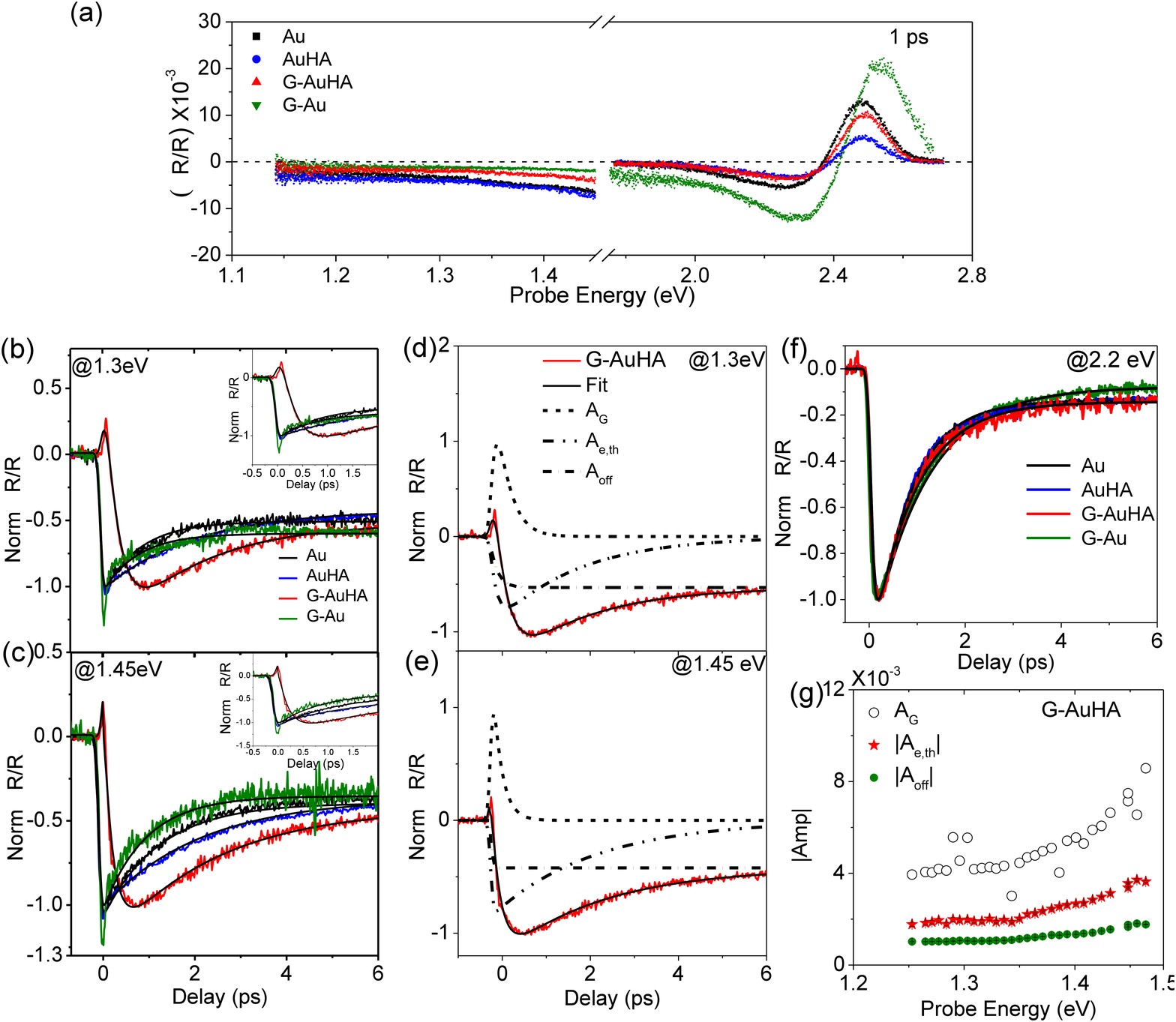}
\caption{(Color Online) (a) 400nm (3.1 eV) pump induced reflection change of Au, AuHA, G-AuHA and G-Au at 1 ps probe delay. Kinetics of Au, AuHA, G-AuHA  and G-Au at probe energies (b) 1.3 eV and (c) 1.45 eV.  Solid lines are fit to the experimental data as discussed in the main text. Inset in (b) and (c) show the zoomed kinetics upto 2 ps delay at probe energies 1.3 eV and 1.45 eV, respectively. (d and e) Fitting components contributing to the total fit of kinetics of G-AuHA in (b) and (c).  (f) Kinetics at probe energy 2.2 eV compared for all the four samples. The plasmon induced signal originating from graphene is not observed in the probe region where plasmonic resonances are not present. (g) Amplitude of the fitting components as a function of probe energy in G-AuHA. } 
\label{fig:Kinetics}
\end{figure*}

\begin{figure*}[h]
\centering
\includegraphics[width=0.9 \linewidth]{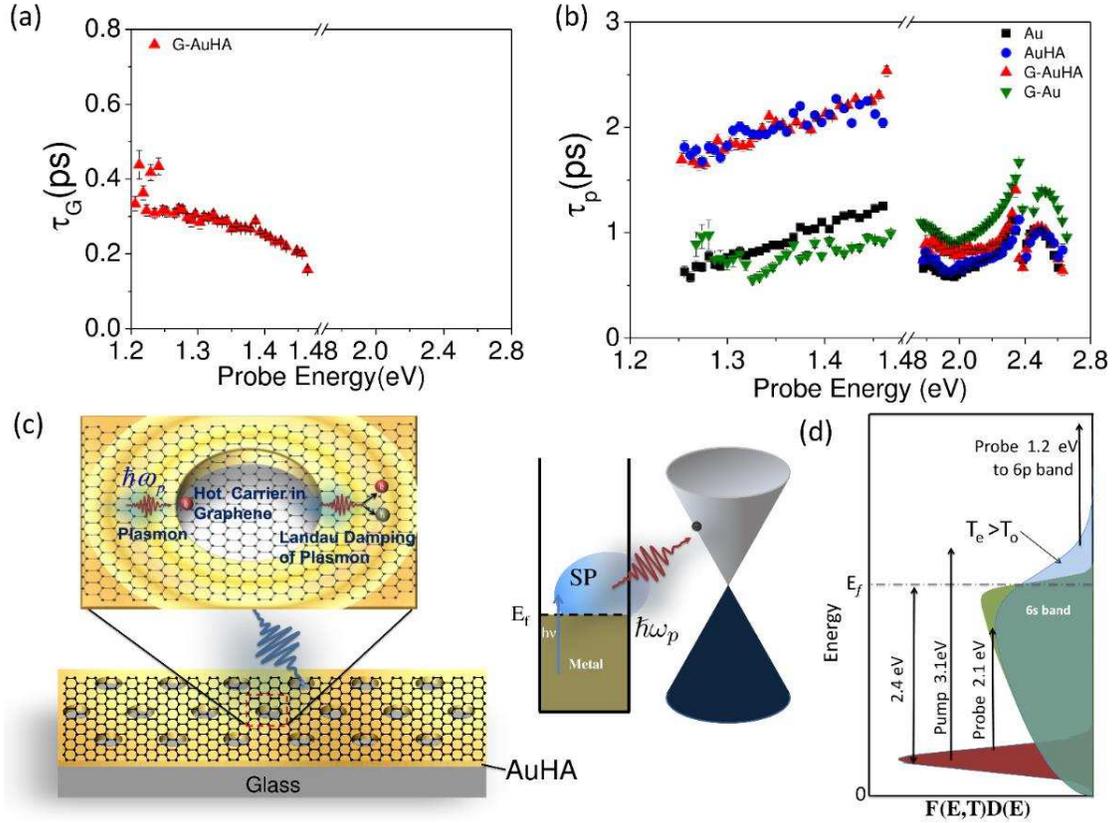}
\caption{(Color Online)(a) Decay time constant $\tau_{G}$ of plasmon induced hot carriers generated in graphene through carrier-optical phonon scattering as a function of probe energy.  (b) Probe photon energy dependence of phonon mediated carrier relaxation time in Au, AuHA, G-AuHA and G-Au. (c) Schematic showing plasmon induced hot electron generation in graphene by charge transfer from Au to graphene and Landau damping of plasmons creating electron- hole pair (d) Illustration qualitatively showing electronic density of states of gold (adopted from Ref. [48]). Excited electrons after thermalization attain an electronic temperature much higher than the lattice temperature. Fermi-Dirac distribution of electrons at an elevated temperature smear carrier occupancies around fermi energy (E$_f$). Probe with energy less than interband transition threshold $\sim$2.4 eV detects holes below fermi energy. }
\label{fig:t_r&t_p}
\end{figure*}

Fig.  \ref{fig:Kinetics}(a) shows transient pump-induced reflectivity change as a function of probe energy in Au, AuHA,  G-Au and G-AuHA at  probe delay of 1 ps  in the spectral window 1.1 to 1.45 eV (NIR) and 1.7 to 2.8 eV (UV-VIS).    The observed pump induced reflectivity change match qualitatively in all the four samples. The shape of the reflectivity change as a function of probe energy depends on the different regions of electronic density of states (DOS) interogated by the probe pulse (as illustrated in Fig.  \ref{fig:t_r&t_p}(d)) \citep{devizis2006ultrafast}. The pump beam at 3.1 eV (higher than the interband transition of gold at 2.4 eV) gets absorbed by the 5d $\rightarrow$ 6s transition and  excites the electron  distribution  out of equilibrium in the conduction band and holes in the 5d band. From the two-photon photoemission (2PPE) studies it has been seen that the holes in the 5d state relax in a very short time, of the order of few tens of femtoseconds, via Auger processes and deliver the excess energy to the 6s band electrons \citep{devizis2006ultrafast, link1999electron, aeschlimann1995femtosecond,  pawlik1997lifetime}. Thus,  the excited 5d holes relax within the pump pulse duration.  The non-Fermi electron distribution, through electron-electron interaction, redistributes its energy and attains a Fermi distribution with temperature T$_e>>$  T$_0$ (initial lattice temperature). The thermalization time of electrons varies from $<$ 100 fs to few hundreds of fs \citep{sun1994femtosecond, eesley1983observation}. The quasi-thermalized electron distribution then loses its energy within a few picoseconds to the lattice through electron-phonon interaction. An equilibrium with the lattice at a slightly higher temperature than T$_0$ is reached afterwards followed by cooling of the hot lattice on a 100 ps time scale  \citep{groeneveld1995femtosecond, sun1994femtosecond}. Following the arguments of Sun et al., the nonthermalized distribution of carriers interacts with phonon bath during the thermalization process. Therefore, exploring the energy exchange mechanism by considering the onset of the electron-electron and electron-phonon relaxation as two seperate regimes is not accurate. However, the energy exchange mechanism can be modeled by taking two separate populations of photoexcited carriers of thermalized and non-thermalized distributions\citep{sun1994femtosecond}. Following \citep{sun1994femtosecond}, we fit the transient reflectivity with the following function convoluted with the Gaussian pulse response:

\begin{eqnarray}
 \frac{\Delta R}{R}\left(\hslash \omega,t \right)&=&\text{H}(t)\lbrace A_{e,_{NT}} e^{-t/\tau_{r'}}  \\&& +  A_{e,th}\left(1-e^{-t/\tau_{r}} \right)e^{-t/\tau_{p}} + A_{G}e^{-t/\tau_{G}} + A_L \rbrace \nonumber
\end{eqnarray}

where H(t) is a heaviside step function. $A_{e,_{NT}}$ and $A_{e,th}$ are the initial transient amplitudes corresponding to non-thermalized and thermalized electronic distributions in Au, respectively. $A_L$  is the offset amplitude due to slow cooling of the lattice.  $\tau_{r}$  and $\tau_{p}$ are the rise time  associated with the thermalization of photoexcited carriers  and the time constant for the decay of the thermalized distribution due to electron-phonon interaction, respectively. $\tau_{r'}=(1/\tau_{r}+1/\tau_{p})^{-1}$ is the decay time of non-thermalized population, taking into account simultaneous interaction of non-thermalized electrons with phonon bath. $A_G$ and $\tau_G$ are the amplitude and decay time of the signal from graphene considered only in the NIR region.   As the amplitude of nonthermalized distribution is shown to be significant only around Fermi energy ($\sim$ 2.4 eV) of gold \citep{sun1994femtosecond}, we include the contribution of $A_{e,_{NT}}$  only near 2.4 eV probe energy. Except for the probe energy $\sim$ 2.4 eV, the rise time $\tau_r$ in all the four samples is limited by the instrument response function ($\sim115$ fs).   The rise time $\tau _r$ is found to be $\sim$250 fs near Fermi energy of gold. The long rise time observed near Fermi energy is  due to state filling effect \citep{sun1994femtosecond}, which blocks the relaxation of electrons close to the Fermi surface while high-energy electrons decay on a much shorter time scales. This is also commensurate  with the Fermi liquid theory which predicts that the relaxation rate varies as (E-E$_f$)$^2$ \citep{sun1994femtosecond}.

Fig.  \ref{fig:Kinetics}(b) and (c) show a comparison of kinetics of photoexcited carriers in G-AuHA with AuHA, G-Au, and Au at two representative central probe energies 1.3 and 1.45 eV in the NIR region. For better S/N ratio and improved fitting the kinetic traces have been averaged with a window of $\Delta E \sim 6 $ meV. The behavior of G-AuHA kinetics in this region compels us to consider an additional exponentially decaying component (A$_{G}e^{-t/\tau_{G}}$)  with positive amplitude A$_{G}$ and decay time $\tau_{G}$ in the fitting function [Eq. (2)].  The positive signal contributing to $\Delta R/R$ is related to  transient carrier dynamics of plasmon induced hot carriers in graphene. The positive sign of the signal originating from graphene is in agreement with the reported ultrafast study of graphene \citep{gilbertson2015plasmon}. No such positive signal in UV-VIS region is observed for G-AuHA (Fig.  \ref{fig:Kinetics}(f)).  It can be seen from the fits that the dynamics is well described by the  function (see Fig. \ref{fig:Kinetics}(d) and (e)). The fit to the experimental data was found to be excellent over the entire probe energies.   Fig.  \ref{fig:Kinetics}(g) shows the amplitudes $A_G$ as function of probe photon energy together with $A_{e,th}$ and $A_G$ in G-AuHA.  The relaxation time constants obtained from the fit are shown in Figs.   \ref{fig:t_r&t_p}(a) and (b).

We see that, as the probe photon energy increases from 1.2 to 1.45 eV,    $\tau_{G}$ decreases from     350 $\pm$ 50 fs to 200 $\pm$ 20 fs. The decay time agrees well with the characteristic decay time of carriers through electron-optical phonon (e-op) scattering in CVD grown graphene \citep{shang2011ultrafast, huang2018high}. The e-op scattering in graphene is mainly due to zone-boundary optical phonons and the decay rate is given by :$\frac{1}{\tau_{G}}=\frac{2 \pi}{\hslash}|\epsilon|W$ \citep{shang2011ultrafast, ando2011zero}. Here $|\epsilon|=E-E_f$ is the electron energy with respect to Fermi energy. $W$ is the dimensionless electron-phonon coupling constant.  The characteristic decay time, its dependence on probe photon energy and e-op coupling constant obtained experimentally confirm that the  positive signal seen in the hybrid G-AuHA structure in the NIR region originates from the carrier dynamics in graphene. As the additional signal is observed only in the G-AuHA sample  in the NIR region having the plasmonic resonances due to the hole array (resonances G1, G2 and G3 in Fig.  \ref{fig:Transm&FDTD_E_prof_plot}(a)), we attribute  the amplitude A$_{G}$  due to transient reflectivity modulated by plasmon induced hot carriers in the graphene with  $\tau_{G}$ as their relaxation time \citep{kumar2016ultrafast, gilbertson2015plasmon, hoggard2013using}.   Schematic in Fig.  \ref{fig:t_r&t_p}(c) illustrates the hot carrier generation mechanism in graphene.  As, the pump beam is not in resonance with the EOT  peaks (G1, G2 and G3),  the induced hot carriers in graphene can not be due to the near field enhanced direct photoexcitation \citep{gilbertson2015plasmon}. The generation of hot carriers in graphene is due to plasmon induced interfacial hot carrier transfer from Au into the graphene through regions in contact with the protruding edges of the nannoholes  having high near field intensities \citep{wu2015efficient}.

In the discussion that follows we examine the relaxation time of thermalized carriers in gold through electron-phonon scattering $\tau_p$ and its dependence on probe energy as shown in  Fig.  \ref{fig:t_r&t_p}(b). We see that $\tau_p$ in the UV-VIS region shows similar  behavior for  Au and AuHA. Interestingly, there is a two fold increase in $\tau_p$ in the NIR region. The presence of graphene in G-Au and G-AuHA has no effect on $\tau_p$ in gold, as expected. The spectral dependence of the different regions probed in the electronic band structure of gold are schematically shown in Fig.  \ref{fig:t_r&t_p}(d)\citep{devizis2006ultrafast}. We will first discuss the dependence of $\tau_p$ in UV-VIS range. $\tau_p$ increases from 0.8 ps at 2.0 eV to 1.2 ps at 2.4 eV and then decreases as probe energy further increases beyond 2.4 eV to 2.8 eV. As crystal structure of gold has only one atom per unit cell, carrier relax through acoustic phonon emission. The decay time varies as $\sim \sqrt{k_BT_L/E}$  where E is the energy of electrons, $k_B$ is Boltzmann constant and $T_L$ is the lattice temperature  \citep{guo2014ultrafast}. The smearing of Fermi-Dirac distribution of the electrons  at the elevated electron temperature T$_e$  creates holes below the Fermi energy.  As a consequence of this the forbidden interband transitions 5d $\rightarrow$ 6s states below the Fermi energy at initial electron temperature T$_e$=T$_0$ are  allowed at T$_e$ $\gg$  T$_0$, whereas  the  states above the Fermi energy are blocked by the phase space filling \citep{devizis2006ultrafast}. Therefore, the probe photons with energy  less than the interband transition threshold (E$_{pr}$ $\sim$ 2.4 eV) detect holes below Fermi energy, whereas greater than the interband transition threshold detect electrons above Fermi energy. Hence, the observed electron-phonon cooling rate is higher when probed at 2.0 eV, increases with probe photon energy up to $E_f\sim$ 2.4 eV and then  decreases  as  probe photon energy moves away from the Fermi energy up to 2.8 eV.

In the NIR region there are three main results : (i) an overall  increase of 1 ps in  $\tau_p$ of AuHA as compared to Au, (ii) the linear increase of $\tau_p$ with probe photon energy and  (iii) as mentioned before, the presence of graphene in G-Au and G-AuHA has no effect on $\tau_p$. As the probe in the NIR region corresponds to intraband transition one would expect photobleach signal ($\Delta R/R$ with a positive sign) due to the thermalized distribution of carrieres near $E_f$. However, as the observed  $\Delta R/R$ signal in this region is negative, a photoinduced absorption is possible  due to the transition from 6s band to 6p band \citep{li2014influence, lasser1981interband}. The probe in this region is excited state absorption by the thermalized carriers above $E_f$ and detects the electron with energy $E=E_{6p} -E_{pr} $ , where $E_{6p}\sim $ 4.32 eV  \citep{li2014influence} is the position of the 6p band and $E_{pr}$ is the probe photon energy. Higher energy probe photons therefore, detect low energy electrons above the Fermi energy. As discussed earlier the thermalized electrons decay through acoustic phonons with decay time-varying as $\sim \sqrt{k_BT_L/E}$,  $\tau_p$ is expected to vary as $\sim \sqrt{k_BT_L/(E_{6p} -E_{pr})}$. Therefore, in this region we see $\tau_p$ increasing with probe photon energy.

We now look at an overall increase of $\tau_p$  by 1 ps in  AuHA in the NIR region. The perforation of gold film with holes can not change the electron-phonon coupling constant of gold and hence, the increase in $\tau_p$ can not be associated to change in shape or filling factor \citep{zhou2016evolution}. However,  the plasmonic behavior of AuHA can alter the relaxation time $\tau_p$. In the NIR region as the  plasmons decay  through Landau damping mechanism, they generate hot electron-hole pairs in AuHA \citep{moskovits2015case, link1999size, sonnichsen2002drastic, link2000shape, bernardi2015theory, groeneveld1995femtosecond}.   The decay time and length of Landau damped plasmons are  $\sim$1-100 fs and $ \sim $1 nm, respectively
 \citep{jia2016interface, brongersma2015plasmon, bernardi2015theory}.  The electromagnetic field intensity due to plasmons being high at the hole edges (Fig.  \ref{fig:Transm&FDTD_E_prof_plot}(c-d)), an initial population of hot electrons confined very close to the edge of the holes  is generated from the plasmon decay within the pulse duration. This gives rise to a   non-uniform distribution  of hot carriers between the holes. The hot carriers diffuse from their high density regions to form spatially uniform distribution of hot elctrons.  The  characteristic hot carrier diffusion time is given by $\tau _d=C_e l^2/2 \kappa_e$, where $C_e$=1.98$\times10^{4}$ Jm$^{-3}$K$^{-1}$ and $\kappa_e$=310 Wm$^{-1}$K$^{-1}$ are electronic specific heat and thermal conductivity of gold, respectively \citep{groeneveld1995femtosecond}. Taking $l$=260 nm as the half of the distance between the edges of two adjacent holes, we get an estimate $\tau_d \sim$2.2 ps. As the diffusion time $\tau_d$ and carrier cooling time $\tau_p$ are comparable, an increase of 1 ps in $\tau_p$ can be possibly attributed to diffusion of hot carriers. Also, as the plasmon-induced hot carriers in graphene are generated through the hole edges (or plasmonic hotspots) in contact with graphene, hot carriers generated are  initially non-uniformly distributed in graphene. The hot carriers in graphene also take finite time to diffuse and form a uniform distribution between the plasmonic hotspots. The  diffusion coefficient of hot carriers in CVD graphene is D$_e$=5.5 $\times10^{3}$ cm$^2$s$^{-1}$ \citep{ruzicka2012spatially}. The characteristic diffusion time of hot carriers  estimated by $\tau _d= l^2/2 D_e$ gives $\tau _d$=57 fs for =250 nm (half of the average distance between hole edges or plasmonic hotspots). The fast diffusion ensures a uniform distribution hot carriers in graphene within the pump pulse duration. Therefore, the presence of graphene has no effect on $\tau_p$ in G-AuHA.    However, a slightly  reduced value of $\tau_p$ is observed in G-Au in comparison to Au film. As graphene is in more contact with Au film it modifies the  decay rate to effective decay rate ($1/\tau_{eff}=1/\tau_p +1/\tau_G$) and therefore results in a slightly faster decay \citep{brida2013ultrafast}.  
 
  A comparison of ultrafast carrier dynamics also helps us in identifying the direct or indirect nature of the plasmon-induced hot carrier generation in graphene. If the transfer pathway of hot carriers in graphene is indirect (due to the near-field interaction of graphene with gold\cite{gilbertson2015plasmon}  or plasmon-induced hot electron transfer (PHET)\cite{wu2015efficient}) the carrier dynamics in graphene must be affected by the diffusion dynamics of hot carriers in AuHA and should show a slower decay dynamics. But, as the time constant $\tau_G$ observed by us in G-AuHA agrees with that reported for CVD Graphene on quartz substrate\citep{shang2011ultrafast} we conclude that the transfer pathway of plasmon-induced hot carriers in graphene in the G-AuHA system is direct. The back transfer of hot carriers to Au is prevented by spatially separating them from the plasmonic hot spots (protruding edges of the holes).

\section{\label{sec:Conclusion} Conclusion}
In conclusion, we have studied photoexcited carrier dynamics in hybrid plasmonic structure of graphene-covered gold hole array to study the graphene-plasmon interaction.   Graphene on the 3D hole array of gold shows strong coupling to the plasmonic modes.  We have shown that plasmon-induced hot carriers can be efficiently generated in graphene. The frequency tuning of EOT resonances in Au hole array (by changing the hole spacing) is a very attractive possibility to inject hot carriers in graphene in spectral regions other than the plasmon resonances in noble metals in the visible region.  Our results have interesting technological implications in hot carrier devices with tunable photon energy. The hot electrons once separated from the plasmonic hot
spots can be utilized in surface catalytic reactions and optoelectronic devices. 
Similar to the reported photocatalysis using visible light overlapping with the surface plasmon resonance of the Au nanoparticles  covered with graphene oxide\citep{kumar2016ultrafast}, we propose that a photocatalysis should be possible by
graphene covered AuHA using photons of energy close to other plasmonic resonances of the
AuHA (0.8 to 1.6eV).

\section*{Acknowledgements}
G.P. and S.N.G. thank Council of Scientific and Industrial Research (CSIR) for Senior Research Fellowship (SRF). A.K.S. and R.K.S. thank the Nanomission Council of the Department of Science and Technology, Government of India, for financial support.


%


\newpage

\end{document}